**Neural origins of self-generated thought: Insights from intracranial electrical stimulation and recordings in humans**

Kieran C. R. Fox[a]

[a] Department of Neurology and Neurological Sciences, Stanford University, Stanford, CA, U.S.A. 94304

*Corresponding author:* Fox, K.C.R. (kcrfox@stanford.edu)

**Abstract**

Functional magnetic resonance imaging (fMRI) has begun to narrow down the neural correlates of self-generated forms of thought, with current evidence pointing toward central roles for the default, frontoparietal, and visual networks. Recent work has linked the actual arising and generation of thoughts more specifically to default network activity, but the limited temporal resolution of fMRI has precluded more detailed conclusions about where in the brain self-created mental content is generated and how this is achieved. Here I argue that the unparalleled spatiotemporal resolution of intracranial electrophysiology (iEEG) in human epilepsy patients can begin to provide answers to questions about the specific neural origins of self-generated thought. I review the extensive body of literature from iEEG studies over the past few decades and show that many studies involving passive recording or direct electrical stimulation throughout the brain all point to the medial temporal lobe as a key site of thought-generation. At the same time, null effects from other brain regions suggest that various other default network hubs, such as the posterior cingulate cortex and inferior parietal lobule, might have only a marginal role (if any) in the self-generation or initiation of mental content like dreams, visual imagery, memories, and prospective simulations. Ultimately, combining a variety of neuroscientific methods that compensate for each other's weaknesses and complement each other's strengths may prove to be the most effective way to understand the brain's remarkable ability to decouple from the immediate environment and generate its own experiences.

**Introduction**

An enormous amount of scientific interest has recently begun to focus on spontaneous and self-generated forms of thought (Andrews-Hanna, Smallwood, & Spreng, 2014; Christoff, 2012; Christoff, Irving, Fox, Spreng, & Andrews-Hanna, 2016; Seli, Risko, Smilek, & Schacter, 2016). As experience sampling studies (see Stawarczyk, this volume) and questionnaires develop an understanding of the associated subjective content (Delamillieure et al., 2010; Diaz et al., 2013; Fox, Nijeboer, Solomonova, Domhoff, & Christoff, 2013; Fox, Thompson, Andrews-Hanna, & Christoff, 2014; Stawarczyk, 2017; Stawarczyk, Majerus, Maj, Van der Linden, & D'Argembeau, 2011), functional neuroimaging research over the past two decades has delineated a rough but increasingly refined picture of general brain recruitment associated with these self-generated forms of thought (Fox, Spreng, Ellamil, Andrews-Hanna, & Christoff, 2015). Throughout this chapter, by 'self-generated thought' I will simply mean mental content that is relatively independent of and unrelated to the immediate sensory environment (Andrews-Hanna et al., 2014; Fox, Andrews-Hanna, & Christoff, 2016); taken broadly, self-generated thought includes mental processes such as stimulus-independent thought (McGuire, Paulesu, Frackowiak, & Frith, 1996), task-unrelated thought (Dumontheil, Gilbert, Frith, & Burgess, 2010), spontaneous thought (Spiers & Maguire, 2006), mind-wandering (Christoff, Gordon, Smallwood, Smith, & Schooler, 2009), creative thinking and insight (Ellamil, Dobson, Beeman, & Christoff, 2012), and dreaming (Fox et al., 2013).

Now that a general picture of the subjective content and neural correlates of self-generated thought is emerging, deeper and more subtle questions are being posed: for instance, whether specific neural correlates are associated with specific self-generated content (Gorgolewski et al., 2014; Tusche, Smallwood, Bernhardt, & Singer, 2014); whether differences in brain morphology are associated with individual tendencies toward certain types of self-generated thinking (Bernhardt et al., 2014; Golchert et al., 2017); what the relationship of self-generated thought is to various psychiatric and neurodegenerative conditions (Christoff et al., 2016); and whether specific neural origin sites of self-generated thought can be identified (Fox et al., 2016). It is this final question that I will focus on throughout this chapter: what brain structures are the primary initiators, drivers, and creators when the brain decouples from its sensory environment and self-generates its own experiences? Is this question even valid? Can there be a specific answer?

      First, I will very briefly review what is known about the broad neural correlates of self-generated thought from functional neuroimaging investigations. These studies point to the primacy of the default network in initiating self-generated thought, but cannot seem to go beyond this level of specificity and offer a more detailed answer due to their inherently poor temporal resolution. Next, I will delve into the relatively smaller (but fast-growing) human intracranial electrophysiology literature to explore the neural origins of self-generated thought in more detail. Starting with the broad set of regions identified by functional neuroimaging as being involved in self-generating thought, I synthesize data from human electrophysiology to hone in on the most likely origin/initiation sites and to tentatively exclude other areas from a primary generative role.

**Functional neuroimaging of self-generated thought: The importance of default network regions**

Non-invasive neuroimaging modalities, particularly functional magnetic resonance imaging (fMRI), have been instrumental in exploring the broad neural correlates and large-scale network dynamics associated with self-generated forms of thought (Beaty, Benedek, Silvia, & Schacter, 2016; Christoff et al., 2016; Ellamil et al., 2016; Fox et al., 2013; Fox et al., 2015; Zabelina & Andrews-Hanna, 2016). A recent quantitative meta-analysis of neuroimaging studies investigating various forms of self-generated thought (including mind-wandering, stimulus-independent thought, and spontaneous mentalizing) found that a wide variety of brain regions appear to be recruited by these processes, including multiple nodes of the default, frontoparietal, and visual networks (Fox et al., 2015; Fig. 1). Although this meta-analysis demonstrated the importance of brain regions and networks beyond the default network to self-generated thought, it was unable to answer more specific questions about the functional roles of each network, or of their temporal primacy in initiating self-generated forms of thinking.

Two specific neuroimaging studies of the origins of self-generated thought are interesting not only for the light they shed on this problem, but also because they exemplify the limitations of fMRI when it comes to identifying specific neural origin sites. Briefly, these studies point to the involvement of default network regions in the initial self-generation of thought, but have not been able to identify which specific components of the network are most important.

*Figure 1.* Meta-analysis of brain areas consistently recruited by self-generated forms of thought.

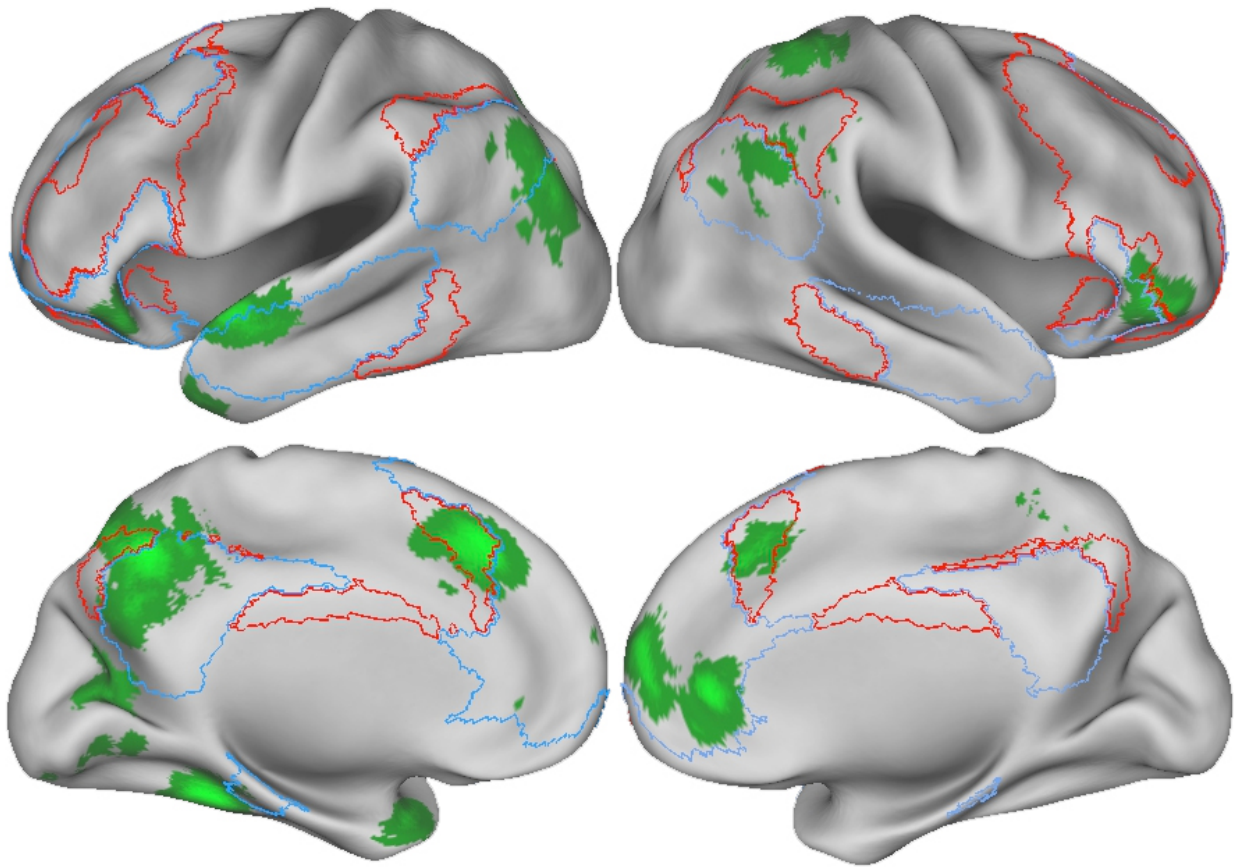

Meta-analytic clusters indicating brain regions consistently recruited across various forms of spontaneous and self-generated thought. Outline of the frontoparietal (red) and default (blue) networks are shown for comparison. Reproduced with permission from Fox et al. (2015).

First, a study by Ellamil and colleagues in 2012 recruited visual artists at a local fine arts university to create visual art while in the scanner using an MRI-compatible drawing tablet and pen (Ellamil et al., 2012). The artists were then asked to reflect on and evaluate the quality of the artwork they had created. The 'generation' phase of artwork creation was contrasted with a control condition where the artists simply traced the pen around the drawing tablet, therefore controlling for motor effects associated with the act of drawing itself. Among the residual activations, associated with the process of generating the artistic ideas themselves, recruitment was observed in the medial temporal lobe (including hippocampus and parahippocampus), inferior and superior parietal lobule, premotor areas, and other regions (Fig. 2). This study helped narrow down the possible regions implicated in self-generation of creative ideas, but nonetheless the widespread recruitment observed made it difficult to determine which regions, if any, were specifically involved in generating creative thoughts, vs. recruited shortly thereafter to participate in this generation process or its communication to other brain regions. (Note that a similar emphasis on the role of the default network in creative generation has been found in a recent study involving participants generating poetry, as opposed to visual artwork (Liu et al., 2015).)

*Figure 2.* Brain regions recruited during self-generation of creative ideas.

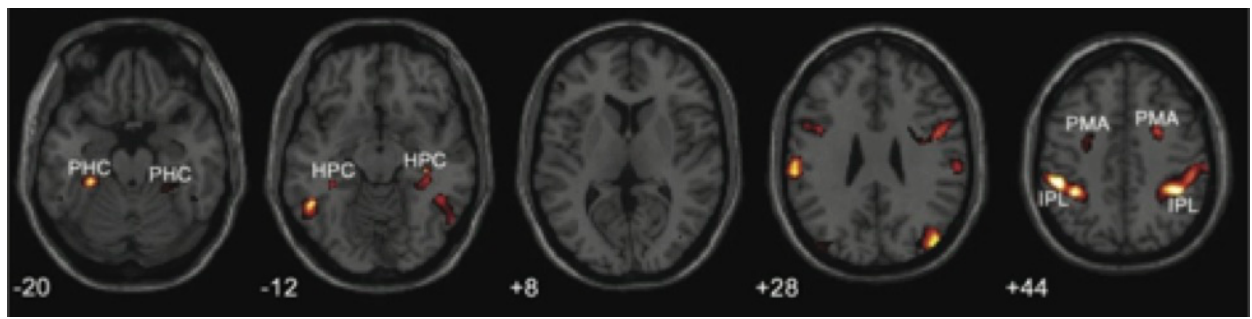

Activations throughout the brain during the generation of visual artwork. Numbers indicate z-coordinates in MNI space. Reproduced with permission from Ellamil et al. (2012). HPC: hippocampus; IPL: inferior parietal lobule; PHC: parahippocampus; PMA: premotor area.

A related study examined brain activation surrounding the spontaneous arising of everyday thoughts when participants' minds wandered while in the MRI scanner (Ellamil et al., 2016). The participants were highly-experienced mindfulness meditation practitioners accustomed to monitoring the arising of distracting thoughts and identifying their contents. While they practiced mindful attention to their breathing, they indicated with a button press when they noticed the arising of a spontaneous, unbidden thought into consciousness. The time-course of brain recruitment before, during, and after this button press was then examined in detail. When these time-courses were explored, it was found that certain brain regions showed peaks of activation slightly *before* the subjective awareness of a thought (as indicated by the button press; see Fig. 3). Various other regions showed peak activations that either coincided with the button press, or followed it (only figures for peak activations *prior* to arising thoughts are shown here; for details of later activation peaks, see the original figures in Ellamil et al. [2016]). Regions showing peak activation antecedent to the arising of thought included the medial temporal lobe bilaterally, the inferior parietal lobule, the posterior cingulate cortex, and others (Fig. 3).

Notably, almost all of these regions are considered components of a broadly defined default network (Buckner, Andrews-Hanna, & Schacter, 2008; Raichle et al., 2001), which (as noted above) is widely agreed to be essential to self-generated and self-referential cognitive activity (Fox et al., 2015; Northoff et al., 2006; Stawarczyk & D'Argembeau, 2015). Beyond this focus on the default network, however, the coarse temporal resolution of fMRI was unable to differentiate between the regions showing antecedent activations, leaving important questions unanswered. Are these various default network or other regions equally involved in generating mental content – or might their dispersed but synchronized activity (functional connectivity) be the explanation? Or are specific areas preferentially involved in generating novel patterns of brain activity corresponding to self-generated mental content?

*Figure 3.* Timecourse of brain regions where activation peaks just prior to awareness of spontaneously arising of thoughts.

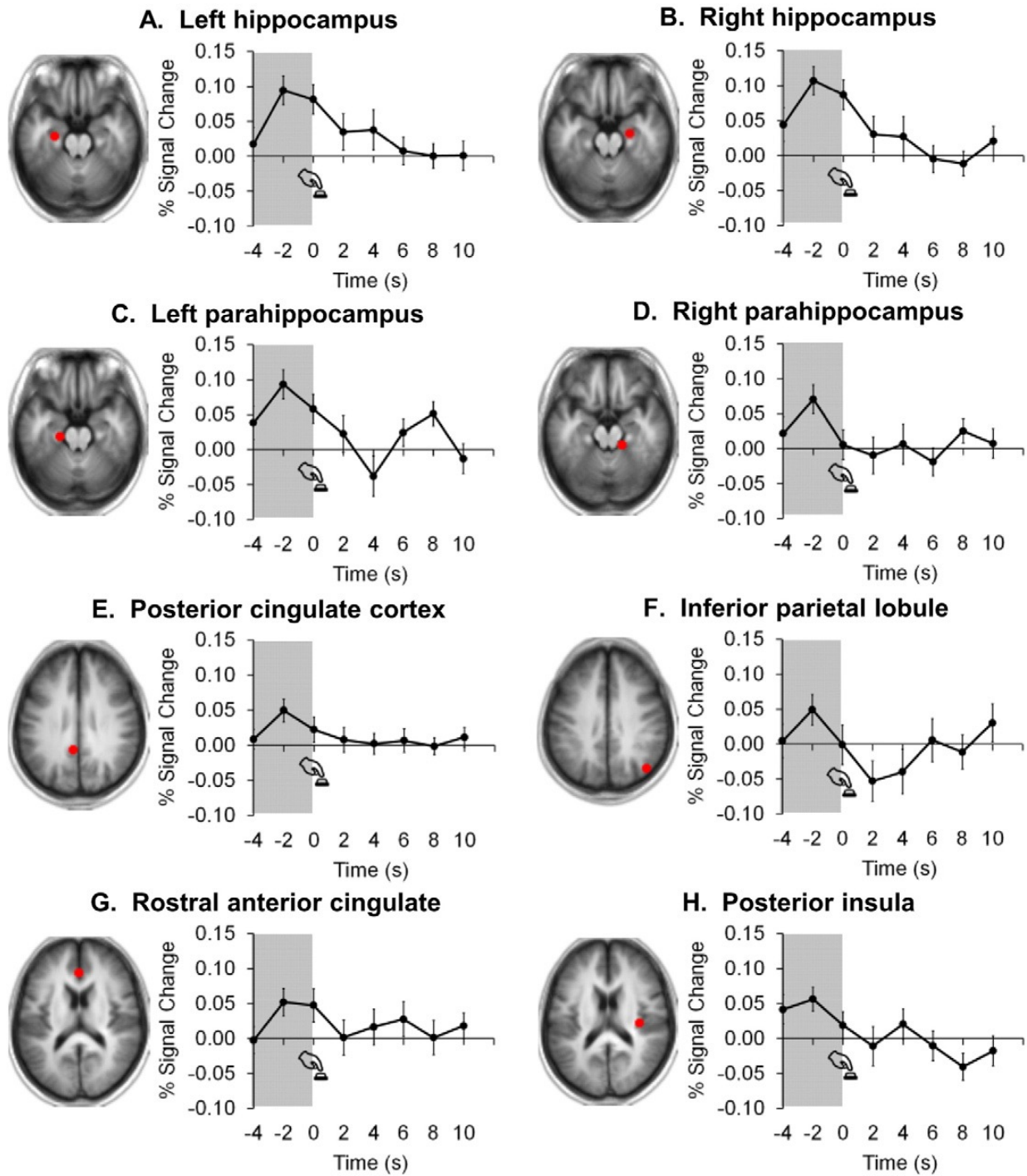

Brain regions where activation peaked prior to the conscious awareness of a spontaneous thought arising (as indicated by the button-press icon). Note that although the results suggest an important role for the medial temporal lobe, the temporal resolution of fMRI could not distinguish these early activations from those in other brain regions, such as the posterior cingulate cortex and rostral anterior cingulate. Reproduced with permission from Ellamil et al. (2016).

**Where are thoughts generated in the brain? Insights from human intracranial electrophysiology**

Direct recording of the electrical activity produced by single neurons and neuronal populations provides unparalleled spatial and temporal resolution (at the scale of single neurons and in the millisecond range) (Fried, Rutishauser, Cerf, & Kreiman, 2014; Suthana & Fried, 2012). By directly recording the brain's electrical activity, be it summation of input signals (potentials in the dendrites and soma), or action potentials carrying a neuron's output signals ('spikes'), many of the pitfalls and uncertainties of the indirect measures used in functional neuroimaging, such as blood-oxygen-level-dependent (BOLD) signal, can be avoided (Logothetis, 2008). An additional advantage is that current can also be 'injected' (passed through electrodes on the cortical surface or at depth in the brain), allowing direct electrical stimulation of tissue throughout the central nervous system. Instead of aiming to evoke brain activity (and then record it) using particular tasks, sensory stimuli, or behaviors, the brain can be directly stimulated and the resulting sensory, cognitive, motor, or emotional effects observed and correlated with the precise site of stimulation (Penfield & Boldrey, 1937; Selimbeyoglu & Parvizi, 2010). Additionally, with multiple simultaneous electrode sites, responses evoked in response to stimulation at a given site can also be investigated (Keller et al., 2014; Matsumoto et al., 2004).

Because of the invasiveness of implanted electrodes, such research cannot be conducted in healthy human participants. In some cases of serious neurological conditions, however, including medication-resistant epilepsy and Parkinson's disease (Bechtereva & Abdullaev, 2000; Lachaux, Rudrauf, & Kahane, 2003), electrical stimulation of the cortical surface or implantation of depth electrodes into deeper cortical and subcortical structures may be indicated in human patients by clinical criteria and protocols (Engel, Moll, Fried, & Ojemann, 2005; Suthana & Fried, 2012). In addition to providing ever-improving clinical benefits for these various conditions (Hill et al., 2017; Spencer, Spencer, Williamson, & Mattson, 1990; Spencer, Spencer, Williamson, & Mattson, 1982), intracranial electrodes provide an unparalleled opportunity to investigate human brain function at very high spatiotemporal resolution.

In many ways, human cognitive electrophysiology remains in its infancy. Nonetheless, the collective results of such investigations in humans have provided unprecedented insights into the understanding of somatosensory and motor systems (Penfield & Boldrey, 1937; Penfield & Welch, 1951), visual perception (Fried, MacDonald, & Wilson, 1997; Kreiman, Koch, & Fried, 2000a; Quiroga, Reddy, Kreiman, Koch, & Fried, 2005), memory (Burke et al., 2014; Gelbard-Sagiv, Mukamel, Harel, Malach, & Fried, 2008; Lega, Jacobs, & Kahana, 2012), spatial mapping and navigation (Ekstrom et al., 2003; Jacobs, Kahana, Ekstrom, Mollison, & Fried, 2010), mathematical cognition (Daitch et al., 2016), visual imagery (Kreiman, Koch, & Fried, 2000b), and even consciousness (Quiroga, Mukamel, Isham, Malach, & Fried, 2008).

This seminal work has only touched on what is possible, however – particularly with respect to higher-order cognitive-affective processes that are difficult, if not impossible, to study in animal models. One such process is the self-generation of mental content by the human brain, the central concern of this Handbook. Although no known human electrophysiology has *directly* explored mental states such as mind-wandering, nonetheless many investigations have explored related phenomena, for instance spontaneous memory recall (Gelbard-Sagiv et al., 2008) and immersive, dream-like experiences (Vignal, Maillard,

McGonigal, & Chauvel, 2007). Moreover, an increasing number of such studies have investigated default network hubs; given this network's acknowledged importance for self-generated thought (Fox et al., 2015), electrophysiological studies of these areas in humans are of great interest. As we shall see, stimulation and recording experiments in default network hubs can also be very informative even when null results are obtained.

In attempting to discern origin sites responsible for creating self-generated mental content, both 'positive' and 'negative' evidence is valuable. By 'positive' evidence we mean that which directly links activity in a given brain area to the subjective experience of self-generated mental content: for instance, electrical stimulation of the medial temporal lobe, as well as spontaneous electrical discharges therein, are both frequently associated with dream-like, hallucinated experiences (Fox et al., 2016; Selimbeyoglu & Parvizi, 2010; Vignal et al., 2007), and spontaneous recall of episodic memories is directly preceded by elevated firing rates in medial temporal lobe neurons (Gelbard-Sagiv et al., 2008). Conversely, 'negative' evidence accrues when stimulation and/or spontaneous discharges fail to result in such subjective experiences, or sometimes any experiences whatsoever (null results): for instance, hundreds of stimulations to the posteromedial cortex in humans have failed to reliably elicit *any* noticeable subjective effects, including self-generated thought (Foster & Parvizi, 2017), despite the importance of this area to the default network (Greicius, Krasnow, Reiss, & Menon, 2003; Raichle et al., 2001; Yeo et al., 2011) and self-generated thought (Fox et al., 2015; Stawarczyk & D'Argembeau, 2015). In the following sections, we summarize what has been learned from human electrical brain stimulation studies about which regions appear to be likely thought-generation or -initiation sites, and which regions do not.

*Table 1.* Summary of human electrophysiology studies demonstrating elicitation of memories, thoughts, or hallucinatory, dream-like experiences.

| Brain Region | Stimulations/ discharges eliciting[a] | Total stimulations/ discharges | Percentage eliciting | References |
|---|---|---|---|---|
| **Temporal Lobe** | | | | |
| Hippocampus | 25 | 46 | 54% | (Bancaud, Brunet-Bourgin, Chauvel, & Halgren, 1994; Fish, Gloor, Quesney, & Oliver, 1993; Halgren, Walter, Cherlow, & Crandall, 1978; Kahane, Hoffmann, Minotti, & Berthoz, 2003; Mulak, Kahane, Hoffmann, Minotti, & Bonaz, 2008; Vignal et al., 2007) |
| Amygdala | 13 | 36 | 36% | (Ferguson et al., 1969; Fish et al., 1993; Halgren et al., 1978; Vignal et al., 2007) |
| Parahippocampal region | 9 | 16 | 56% | (Feindel & Penfield, 1954; Penfield & Perot, 1963; Vignal et al., 2007) |
| Temporopolar cortex | 5 | 11 | 45% | (Bancaud et al., 1994; Halgren et al., 1978; Mulak et al., 2008; Ostrowsky, Desestret, Ryvlin, Coste, & Mauguière, 2002; Penfield & Perot, 1963) |
| Inferior temporal gyrus | 1 | 21 | 5% | (Penfield & Perot, 1963) |
| Middle temporal gyrus | 7 | 42 | 17% | (Kahane et al., 2003; Mullan & Penfield, 1959; Penfield, 1958; Penfield & Perot, 1963) |
| Superior temporal gyrus | 24 | 99 | 24% | (Morris, Luders, Lesser, Dinner, & Hahn, 1984; Mullan & Penfield, 1959; Penfield & Perot, 1963) |
| Temporo-occipital junction | 4 | 17 | 24% | (Lee, Hong, Seo, Tae, & Hong, 2000; Morris et al., 1984; Penfield & Perot, 1963) |
| **Frontal Lobe** | | | | |
| Inferior frontal gyrus | 1 | 7 | 14% | (Blanke, Landis, & Seeck, 2000) |
| Middle frontal gyrus | 2 | 8 | 25% | (Blanke, Landis, et al., 2000) |
| Orbitofrontal cortex | 1 | 4 | 25% | (Mahl, Rothenberg, Delgado, & Hamlin, 1964) |
| Supplementary motor area | 1 | 6 | 17% | (Beauvais, Biraben, Seigneuret, Saïkali, & Scarabin, 2005) |
| **Parietal Lobe** | | | | |
| Inferior parietal lobule | 2 | 42 | 5% | (Blanke, Perrig, Thut, Landis, & Seeck, 2000; Schulz, Woermann, & Ebner, 2007) |
| Posteromedial cortex (including posterior cingulate cortex) | 0 | 248 | 0% | (Foster & Parvizi, 2017) |

Based on data in Supplementary Table 1 in the comprehensive review conducted by Selimbeyoglu & Parvizi (2010). Updated from a previously published table (Fox et al., 2016). Data for brain areas with ≥10 stimulations/discharges reported in the literature are visualized in Fig. 4. The null effects in posteromedial cortex are included because of its inherent interest as a major default network hub.

*Figure 4*. Preferential involvement of medial temporal lobe structures and temporopolar cortex in electrophysiological stimulations (or spontaneous discharges) eliciting memories, thoughts, or hallucinatory, dream-like experiences.

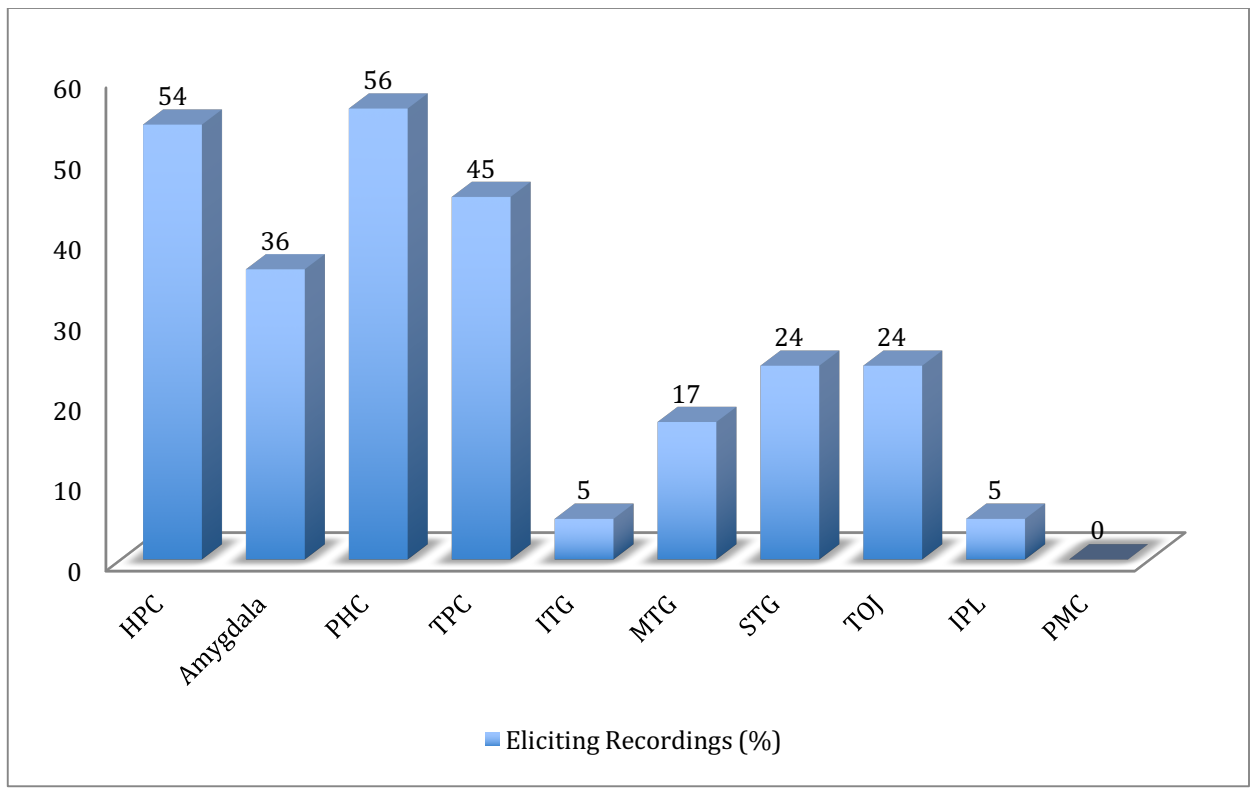

Percentage of stimulations or spontaneous discharges that elicited a first-person experience of memories, thoughts, or hallucinatory, dream-like experiences, based on more than 100 independent investigations. Not shown are data for hundreds of other stimulations throughout the brain, for which no such thought- or dream-like experiences have ever been reported. Only brain areas with ≥10 stimulations or discharges reported in the literature are visualized. Drawn from data in our Table 1, based on data in Supplementary Table 1 of the comprehensive review of Selimbeyoglu & Parvizi (2010); updated and modified based on our previously published figure (Fox et al., 2016). HPC: hippocampus; IPL: inferior parietal lobule; ITG: inferior temporal gyrus; MTG: middle temporal gyrus; PHC: parahippocampal cortex; PMC: posteromedial cortex; STG: superior temporal gyrus; TOJ: temporo-occipital junction; TPC: temporopolar cortex.

*Positive evidence: The importance of the medial temporal lobe and temporopolar cortex in initiating self-generated thought*

**Medial temporal lobe.** The most substantial evidence to date points to the medial temporal lobe as a primary (if not the only) origin site for self-generating mental content of various kinds. Electrical brain stimulation to both the hippocampus (Bancaud et al., 1994; Fish et al., 1993; Halgren et al., 1978; Kahane et al., 2003; Mulak et al., 2008; Vignal et al., 2007) and parahippocampal region (Feindel & Penfield, 1954; Penfield & Perot, 1963; Vignal et al., 2007) elicit subjective experiences akin to self-generated thought more than half of the time (according to reports in the existing literature) (Table 1; Fig. 4) – considerably more often than any other area studied to date, with the possible exception of the temporopolar cortex (discussed in the next section). These experiences include memory recall (Penfield & Perot, 1963), visual hallucinations (Halgren et al., 1978; Kahane et al., 2003; Penfield & Perot, 1963; Vignal et al., 2007), and dreaming (Halgren et al., 1978; Vignal et al., 2007).

Stimulation of the amygdala also elicits such subjective experiences in about one third of cases (Table 1; Fig. 4). This fact is intriguing because the amygdala is the only subcortical structure reported to regularly elicit self-generated thought (Fig. 4). Stimulation of other subcortical structures, such as the thalamus, globus pallidus, and subthalamic nucleus results in a variety of interesting effects, but none of them bear much resemblance to self-generated thought the way we have defined it (Selimbeyoglu & Parvizi, 2010). Similar to the cortical medial temporal lobe structures, stimulation of amygdala can elicit long-term memory recall (Fish et al., 1993; Vignal et al., 2007), out-of-body experiences (Vignal et al., 2007), visual hallucinations (Fish et al., 1993), and dreams (Vignal et al., 2007).

**Temporopolar cortex.** Although reports of stimulation to the temporopolar cortex are relatively rare, about half of these stimulations have resulted in subjective experiences resembling self-generated thought (Bancaud et al., 1994; Halgren et al., 1978; Mulak et al., 2008; Ostrowsky et al., 2002; Penfield & Perot, 1963). As with the medial temporal lobe, stimulation can elicit visual hallucinations (Bancaud et al., 1994; Halgren et al., 1978; Penfield & Perot, 1963), memories (Bancaud et al., 1994), and other thoughts (Ostrowsky et al., 2002). Although this high rate of elicitation of self-generated mental content is suggestive, far too few stimulations have been conducted for any firm conclusions to be drawn. Further stimulations of temporopolar cortex alongside reports of subjective experiences would be highly valuable to a deeper understanding of its role in self-generated thought (see also Discussion, below).

**Other regions.** There is also limited evidence for the importance of other regions, most notably the middle temporal gyrus (Kahane et al., 2003; Mullan & Penfield, 1959; Penfield, 1958; Penfield & Perot, 1963), superior temporal gyrus (Morris et al., 1984; Mullan & Penfield, 1959; Penfield & Perot, 1963), and temporo-occipital junction (Lee et al., 2000; Morris et al., 1984; Penfield & Perot, 1963). Isolated reports of experiences such as complex visual hallucinations elicited by stimulation of the frontal lobe are also interesting (Blanke, Landis, et al., 2000; Blanke, Perrig, et al., 2000). Because the evidence in these cases is much more marginal, we do not discuss them further here, but suffice to say that the book is hardly closed on the involvement of other brain areas in self-generating mental content. Further details of these isolated reports can be found in Table 1 and Figure 4.

Also important is the general dearth of stimulation to subcortical areas reported in the literature: more information about subjective effects of stimulation to the brainstem, cerebellum, and other subcortical regions would be a welcome addition to the literature. For a comprehensive review of subjective effects elicited by stimulation of subcortical structures, see the review by Selimbeyoglu and Pariviz (2010).

*Negative evidence: Marginal roles for posteromedial cortex and inferior parietal lobule in initiating self-generated thought*

**Posteromedial cortex.** The broad area designated as posteromedial cortex (Parvizi, Van Hoesen, Buckwalter, & Damasio, 2006), including posterior cingulate cortex (BA 23/31), precuneus (BA 7), and retrosplenial cortex (BA 29/30), is another likely candidate area for initial generation of thought content. Posteromedial cortex is hypothesized to play a key role in initiating or facilitating memory recall (Shannon & Buckner, 2004), and memory recall and recombination represent a considerable proportion of self-generated thought (Andrews-Hanna, Reidler, Huang, & Buckner, 2010; Fox et al., 2013; Stawarczyk, 2017). The posterior cingulate cortex is consistently recruited during various forms of self-generated thought (Fox et al., 2015; Stawarczyk & D'Argembeau, 2015), and there is evidence from fMRI that activity there peaks in the 2 s time window during which spontaneous thoughts appear to be arising ((Ellamil et al., 2016); Fig. 3). Moreover, intracranial electrophysiological studies in humans have shown that distinct neuronal populations in the posterior cingulate show elevated high-gamma-band activity both at rest and during self-referential thinking (Dastjerdi et al., 2011).

Despite these promising features, however, the evidence to date from intracranial electrophysiological investigations in humans argues against any causal role for the posterior cingulate cortex, as well as adjacent retrosplenial cortex, in generating or initiating mental content. The most comprehensive study of this region to date was published only recently (Foster & Parvizi, 2017). Exhaustively cataloguing more than 800 electrical stimulations throughout medial posterior brain regions across 25 epilepsy patients, Foster and Parvizi (2017) found that stimulation to centrally-located posteromedial brain regions *never* yielded subjective experiences or disturbances of any kind, unless stimulations were executed on dorsal or ventral border regions (Fig. 5).

A possible exception to these results should be mentioned: a single-patient case study recently reported that electrical stimulation of the posterior cingulate cortex disconnects consciousness from the external environment and results in subjective experiences of 'dreaming' (Herbet et al., 2014). If correct, these findings could suggest that posterior cingulate cortex, too, could be considered a powerful initiatory/generative site for self-generated thought (the authors' interpretation of their findings, conversely, is that electrical stimulation 'disrupts' the posterior cingulate cortex connectivity which underlies conscious attention to the external world, thus making room for internally-generated, dream-like experiences). There are important limitations to this study, however, that render either interpretation unlikely. The central concern is that electrical brain stimulation was delivered not to the grey matter of the posterior cingulate cortex, but instead to the underlying white matter (in the posterior part of the cingulum; see their Fig. 2a). The authors' interpretations of their findings, which focus on discussion of the posterior cingulate cortex, are problematic given that stimulation was in fact delivered to a white matter pathway known to project widely throughout the brain (Schmahmann et al., 2007; Wakana, Jiang, Nagae-Poetscher, Van Zijl, & Mori, 2004). Indeed, the cingulum is especially problematic in this respect because two of its main projections are to medial prefrontal cortex and medial temporal lobe, both of which are strongly recruited during waking mind-wandering or 'daydreaming' (Ellamil et al., 2016; Fox et al., 2015) and REM sleep, where dreaming usually occurs (Domhoff & Fox, 2015; Fox et al., 2013). Although the authors argue that the partial excision of the cingulum in their patient makes downstream stimulation an unlikely explanation (Herbet et al., 2014), nonetheless their results are very reminiscent of studies that delivered current directly to the medial temporal lobe and elicited dream-like experiences and dreamy states (Bancaud et al., 1994; Halgren et al., 1978; Penfield & Perot, 1963; Vignal et al., 2007). To summarize, the authors interpret their findings as related to posterior cingulate cortex function, whereas their paradigm involved stimulation of a major fiber pathway also connected to medial prefrontal cortex and medial temporal lobe, both of which appear to be directly involved in the creation of dreamy states and other forms of self-generated thought (Domhoff & Fox, 2015; Ellamil et al., 2016; Fox et al., 2013; Fox et al., 2015). Moreover, their localization of the stimulation electrodes appears to have been based on visual assessment during surgery, as opposed to a more reliable localization based on CT or MRI scans in patients. Given these limitations, this single-subject case study is insufficient to outweigh the results of some 25 patients where

Forthcoming chapter in *The Oxford Handbook of Spontaneous Thought*    19

hundreds of stimulations to the posteromedial cortex yielded no such effects (Foster & Parvizi, 2017).

*Figure 5.* Null effects of electrical brain stimulation in the posteromedial cortex default network hub.

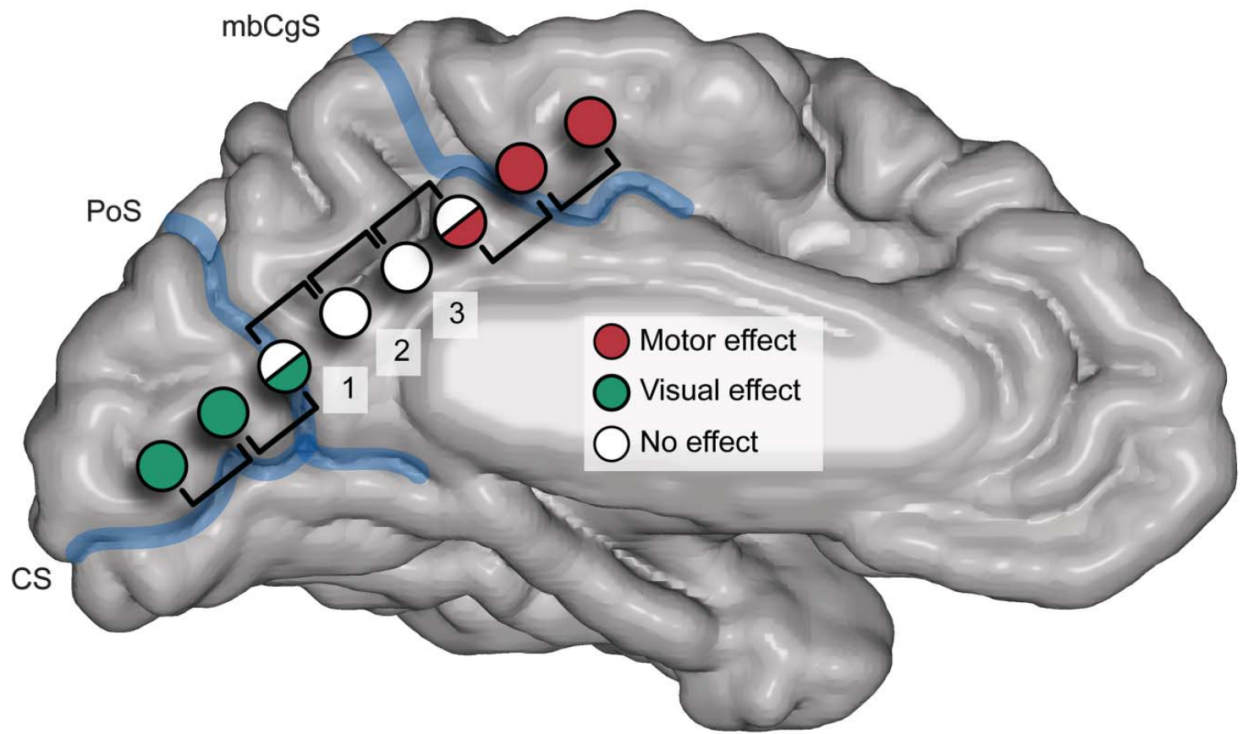

Summary figure showing subjective effects produced by electrical brain stimulation of various regions of the medial posterior portions of cerebral cortex. More dorsal stimulations preferentially evoke motor effects (red circles), and more ventral stimulations largely evoke visual effects (green circles). Some 248 stimulations of more central regions, however (white circles), corresponding closely to a major hub of the default network and overlapping with numerous regions known to be recruited by self-generated thought (cf. Fig. 1), resulted in no discernible subjective effects of any kind. Reproduced with permission from Foster & Parvizi (2017).

**Inferior parietal lobule.** Despite its consistent recruitment during various forms of self-generated thought ((Andrews-Hanna et al., 2014; Fox et al., 2015); Fig. 1) and its undisputed role in the default network (Buckner et al., 2008; Yeo et al., 2011), the inferior parietal lobule does not appear to be a critical origin site for self-generated thought. Two studies have undertaken fairly extensive stimulations to the inferior parietal lobule (Blanke, Perrig, et al., 2000; Schulz et al., 2007); of more than 40 stimulations to this region, only two (5%) elicited some subjective experience reminiscent of self-generated thought: whereas most stimulations elicited simple sensorimotor phenomena, one elicited visual hallucinations (Schulz et al., 2007) and another a self-described 'out-of-body' experience reminiscent of dreaming (Blanke, Perrig, et al., 2000). Although these exceptions are intriguing, given that some 95% of stimulations to inferior parietal lobule have yielded no such effects (Fig. 4), the safest conclusion at present is that it should not be considered as a primary thought generation center. It is important to note however that 'inferior parietal lobule' designates a large swathe of cortex where several networks meet one another (Yeo et al., 2011), and clearly not all of the stimulations reported here would fall within the boundaries of the default network. A more fine-grained approach might find that stimulation specifically to the default network subsection of the inferior parietal lobule results in a higher rates of elicitation of self-generated thought.

**Discussion and future directions**

*Caveat: The varieties of electrical brain stimulation*

An important caveat to all the aforementioned results and conclusions is that electrical brain stimulation is not uniform: voltage, frequency, duration, and other parameters can differ markedly across experiments, and sometimes even within the same patient. Despite this wide variety of stimulation parameters, there is also a fair degree of consistency in some respects – for instance, the majority of studies used 50-60 Hz stimulation – but other factors, such as the current and duration of stimulation, are more variable (Selimbeyoglu & Parvizi, 2010). These facts are most relevant to our discussion of posteromedial cortex, where null results appear to be the rule (Foster & Parvizi, 2017). The apparent 'silence' of posteromedial cortex may not be definitive: varying stimulation parameters (e.g., increasing its duration and/or strength) in future studies could conceivably lead to different (and positive) results in both posteromedial cortex (Foster & Parvizi, 2017) and other brain areas.

*What is the role of medial prefrontal cortex?*

A major lacuna in our knowledge is the role played by the medial prefrontal cortex, which has been relatively rarely investigated in terms of subjective effects in human intracranial patients (Selimbeyoglu & Parvizi, 2010). Medial prefrontal cortex is one of the major hubs of the default network (Buckner et al., 2008; Raichle et al., 2001; Yeo et al., 2011) and is strongly recruited by essentially every form of self-generated thought examined to date, including mind-wandering/daydreaming (Fox et al., 2015), creative thinking (Liu et al., 2015), and dreaming (Domhoff & Fox, 2015; Fox et al., 2013). Future studies elucidating the role of this region are therefore critical to a more complete understanding of the neural origins of self-generated thought.

*The enigmatic temporopolar cortex*

The high rate of self-generated thought elicited by stimulation of the temporal pole (Fig. 4) is an intriguing finding that should be followed up with further research. Temporopolar cortex remains a relatively little-studied and poorly-understood region, as reflected in the title of a recent comprehensive review: 'the enigmatic temporal pole' (Olson, Plotzker, & Ezzyat, 2007). However, anatomists have noted that its pattern of connectivity with other brain regions is strikingly similar to that of the amygdala – another region that appears central to thought generation (Fig. 4; Fox et al., 2016). This connectivity includes dense interconnections with the orbitofrontal cortex, amygdala, and insula (Gloor, 1997; Kondo, Saleem, & Price, 2003; Olson et al., 2007; Stefanacci, Suzuki, & Amaral, 1996), often leading to its grouping with medial temporal lobe and orbitofrontal/medial prefrontal cortex as a limbic or paralimbic area (Mesulam, 2000; Olson et al., 2007). Large-scale investigations of intrinsic resting state functional connectivity substantially agree with this conclusion, grouping temporopolar cortex with medial temporal areas and orbitofrontal/medial prefrontal cortex in a putative 'limbic' network (Yeo et al., 2011).

Anatomical, electrophysiological, and self-report data therefore all point toward temporopolar cortex as functionally and anatomically related to deeper medial temporal lobe structures. It should therefore come as little surprise that stimulation of this region likewise elicits various kinds of self-generated thought. Task-based functional neuroimaging further supports these links: temporopolar cortex recruitment has been reported in numerous independent fMRI and PET investigations of self-generated thought (Christoff et al., 2009; Christoff, Ream, & Gabrieli, 2004; Dumontheil et al., 2010; Ellamil et al., 2016; McGuire et al., 1996), as well as our recent meta-analysis of these forms of cognition (Fox et al., 2015; Fig. 1).

More detailed explanations of the temporopolar cortex's function have also been proposed: for instance, a large number of fMRI studies investigating mentalizing and 'theory of mind' tasks have observed recruitment of temporopolar cortex (Binder, Desai, Graves, & Conant, 2009; Olson et al., 2007), prompting the recent proposal that this area may flexibly couple with other default network components to facilitate these and other social cognitive processes (Spreng & Andrews-Hanna, 2015). One possibility, therefore, is a role in mediating the large amount of social cognition known to take place during self-generated thought (Diaz et al., 2013; Klinger, 2008). Given the poor contemporary understanding of temporopolar cortex, however, and other proposed functional roles (for instance, binding visceral-affective assessments with highly-processed perceptual information [Olson et al., 2007]), the door should be kept open to any number of other possibilities.

*What is special about medial temporal lobe?*

Although a detailed discussion is beyond the scope of this chapter, some features of medial temporal lobe circuitry may help in explaining the important role played by this region (Fox et al., 2016). Whereas typical neocortical circuitry involves a preponderance of local (short-distance) and a corresponding paucity of long-distance connections (Douglas & Martin, 2004; Markram et al., 2004; Thomson & Bannister, 2003), hippocampal neurons appear equally likely to contact near and distant neighbors (Buzsaki, 2006; Li, Somogyi, Tepper, & Buzsaki, 1992; Li, Somogyi, Ylinen, & Buzsaki, 1994). Given that single neurons in medial temporal lobe can encode very high-level, invariant representations of the world (for instance, highly specific famous faces or landmarks; (Quiroga et al., 2005), this specialized and densely interconnected microcircuitry might provide a flexible substrate for encoding novel and arbitrary associations between one percept or idea and another. In the hippocampus's role as a spatial map, this capacity is thought to be critical in that it allows the mapping of 'anything' to 'anywhere' (Eichenbaum, Dudchenko, Wood, Shapiro, & Tanila, 1999; Ekstrom et al., 2003; O'Keefe & Nadel, 1978) – in principle, any known object or person could be set in any known space in the world, and the medial temporal lobe needs to be able not only to represent these arbitrary associations of object and place in perception, but also consolidate them to long-term memory.

If a specialized microcircuitry indeed evolved in the hippocampus allowing for the arbitrary combination of neural activity encoding high-level percepts (in principle allowing the matching of any object or person to any spatiotemporal locus), this capacity could also be 'hijacked' via a process of exaptation – i.e., the recruitment or involvement of a given structure in some function other than that for which it originally evolved (Gould, 1991; Gould & Vrba, 1982) – and utilized for the self-generation of novel/arbitrary combinations of memory traces. From the subjective perspective, the result of this process would be experienced as the spontaneous arising of thoughts, sudden insights and creative ideas, visual imagery and simulations, and even entirely *sui generis* spatiotemporal landscapes during what we call 'dreaming' (Domhoff & Fox, 2015; Fox et al., 2013; Windt, 2010).

On this view, specific patterns of activity initiated in the medial temporal lobe would then recruit (or spread to) regions throughout the brain, likely in a content-dependent manner. There is some preliminary evidence for this kind of association between medial temporal lobe activity and that of other brain regions during self-generated thought. For instance, in an fMRI study of mind-wandering, both overall functional connectivity, as well as fluctuations (variability) in functional connectivity, between the medial temporal lobe subsystem of the default network and the posterior cingulate cortex tracked self-reports of daydreaming frequency (Kucyi & Davis, 2014). On a finer timescale, a study using intracranial EEG investigating spontaneous memory recall found that medial temporal lobe structures showed the earliest peaks in high γ-band activity, whereas γ-band peaks were observed slightly later in other areas throughout the temporal, parietal, and frontal lobes (Burke et al., 2014). The medial temporal lobe was not the only area to show high-frequency activity peaking prior to spontaneous recall, but it was the only area where this high γ-band activity successfully predicted subsequent memory recall, highlighting its importance both in initiating and predicting the success of spontaneous memory recall (Burke et al., 2014). These findings suggest that spontaneous memory recall might be primarily initiated or generated in the medial temporal lobe, followed by a slightly delayed but much more widespread recruitment of other regions throughout the brain – similar to our hypothesis for spontaneous self-generation of mental content other than memory.

*Conclusions and future directions*

Functional neuroimaging has highlighted the importance of several brain networks to self-generated thought, most notably the default, visual, and frontoparietal networks (Fox et al., 2015), and narrowed down the most likely initiation/generation sites to somewhere within the default network (Ellamil et al., 2012; Ellamil et al., 2016). Ultimately, however, all existing noninvasive neuroimaging modalities lack the spatiotemporal resolution to answer subtle questions about *where* in the brain self-created content is actually generated, as well as how and where this initial self-generated activity subsequently spreads (Fox et al., 2016). Human intracranial electrophysiology, despite being confined to clinical contexts, has helped to again narrow our focus, pointing toward medial temporal lobe and temporopolar cortex as especially relevant to thought generation, while simultaneously pointing to an only marginal role (if any) for other default network hubs, including the posterior cingulate cortex (Foster & Parvizi, 2017) and inferior parietal lobule (Fox et al., 2016). Future work will need to corroborate, contest, and further refine these coarse generalizations, and understudied but potentially important regions, such as medial prefrontal cortex, will need to be more heavily investigated.

responses to electrical brain stimulation in patients with temporal lobe epilepsy. *Psychosomatic Medicine, 31*(6), 479-498.

Fish, D., Gloor, P., Quesney, F., & Oliver, A. (1993). Clinical responses to electrical brain stimulation of the temporal and frontal lobes in patients with epilepsy Pathophysiological implications. *Brain, 116*(2), 397-414.

Foster, B. L., & Parvizi, J. (2017). Direct cortical stimulation of human posteromedial cortex. *Neurology, 88*(7), 685-691.

Fox, K. C. R., Andrews-Hanna, J. R., & Christoff, K. (2016). The neurobiology of self-generated thought from cells to systems: Integrating evidence from lesion studies, human intracranial electrophysiology, neurochemistry, and neuroendocrinology. *Neuroscience, 335*, 134-150.

Fox, K. C. R., Nijeboer, S., Solomonova, E., Domhoff, G. W., & Christoff, K. (2013). Dreaming as mind wandering: Evidence from functional neuroimaging and first-person content reports. *Front Hum Neurosci, 7*, 412. doi:10.3389/fnhum.2013.00412

Fox, K. C. R., Spreng, R. N., Ellamil, M., Andrews-Hanna, J. R., & Christoff, K. (2015). The wandering brain: Meta-analysis of functional neuroimaging studies of mind-wandering and related spontaneous thought processes. *Neuroimage, 111*, 611-621.

Fox, K. C. R., Thompson, E., Andrews-Hanna, J. R., & Christoff, K. (2014). Is thinking really aversive? A commentary on Wilson et al.'s "Just think: The challenges of the disengaged mind". *Frontiers in Psychology: Cognition*.

Fried, I., MacDonald, K. A., & Wilson, C. L. (1997). Single neuron activity in human hippocampus and amygdala during recognition of faces and objects. *Neuron, 18*(5), 753-765.

Fried, I., Rutishauser, U., Cerf, M., & Kreiman, G. (2014). *Single Neuron Studies of the Human Brain: Probing Cognition*: MIT Press.

Gelbard-Sagiv, H., Mukamel, R., Harel, M., Malach, R., & Fried, I. (2008). Internally generated reactivation of single neurons in human hippocampus during free recall. *Science, 322*(5898), 96-101.

Gloor, P. (1997). *The temporal lobe and limbic system*: Oxford University Press, USA.

Golchert, J., Smallwood, J., Jefferies, E., Seli, P., Huntenburg, J. M., Liem, F., . . . Villringer, A. (2017). Individual variation in intentionality in the mind-wandering state is reflected in the integration of the default-mode, fronto-parietal, and limbic networks. *Neuroimage, 146*, 226-235.

Gorgolewski, K. J., Lurie, D., Urchs, S., Kipping, J. A., Craddock, R. C., Milham, M. P., . . . Smallwood, J. (2014). A correspondence between individual differences in the brain's intrinsic functional architecture and the content and form of self-generated thoughts. *PLoS ONE, 9*(5), e97176.

Gould, S. J. (1991). Exaptation: A crucial tool for an evolutionary psychology. *Journal of social issues, 47*(3), 43-65.

Gould, S. J., & Vrba, E. S. (1982). Exaptation—a missing term in the science of form. *Paleobiology, 8*(01), 4-15.

Greicius, M. D., Krasnow, B., Reiss, A. L., & Menon, V. (2003). Functional connectivity in the resting brain: a network analysis of the default mode hypothesis. *Proceedings of the National Academy of Sciences, 100*(1), 253-258.